\documentstyle[12pt]{article}
\font \greekb=cmmib10 scaled \magstep1
\newcommand{\taub}{\mbox{\greekb \char 28}}
\newcommand{\pib}{\mbox{\greekb \char 25}}
\newcommand{\thetab}{\mbox{\greekb \char 18}}
\newcommand{\phib}{\mbox{\greekb \char 30}}
\newcommand{\sigmab}{\mbox{\greekb \char 27}}
\begin{document}

\vspace*{1.0 cm}

\begin{center}
{\Large \bf Double Symmetries in Field Theories} \\

\vspace{1.0 cm}

{\large L.M.Slad \footnote{E-mail: slad@theory.npi.msu.su}} \\

\vspace{0.4 cm}

{\it D.V.Skobeltsyn Institute of Nuclear Physics, \\
Moscow State University, Moscow 119899}
\end{center}

\vspace{0.5 cm}

\centerline{\bf Abstract}

In the paper a concept of a double symmetry is introduced, and its 
qualitative characteristics and rigorous definitions are given. We describe
two ways to construct the double-symmetric field theories and present an 
example demonstrating the high efficiency of one of them. In noting the 
existing double-symmetric theories we draw attention at a dual status 
of the group $SU(2)_{L} \otimes SU(2)_{R}$ as a secondary symmetry group, 
and in this connexion we briefly discuss logically possible aspects of the 
$P$-violation in weak interactions.

\newpage

{\bf 1. Qualitative characteristics of the double symmetry}

\vspace{0.2 cm}

Extension of symmetry approaches to the field theories construction is of an
ordinary practice in physics. In the present paper we propose some
generalization of already existing approaches which consist in constructing a
new group ${\cal G}_{T}$, called a double symmetry group, on the basis of the
given global group $G$ (the primary symmetry group) and some its 
representation $T$. This construction is carried out in the framework of 
group transformations in field-vectors space. Namely, the transformations of 
the double symmetry group consist of transformations of the primary and 
secondary symmetries, realized in the same field-vectors space of some 
representation $S$ of the group $G$. The secondary symmetry transformations, 
which can be global as well as local, have two features. Firstly, the 
parameters of these transformations belong to the space of the representation 
$T$ of the group $G$. Secondly, the secondary symmetry transformations do not 
violate the primary symmetry.

As a rule, the secondary symmetry transformations couple different 
irreducible representations, into which the representations $S$ of the group 
$G$ is decomposed, removing or reducing the arbitrariness of a field theory 
allowed by its invariance in respect to $G$. If one or another secondary 
symmetry corresponds to the real physical world, then this correspondence is 
almost inevitably concealed by its spontaneous breaking.

We restrict ourselves with formulations of the global symmetries. The
knowledge of the secondary symmetry group allows to introduce the relevant 
local transformations and gauge fields in the standard way, if there is any 
need of this.

\vspace{0.3 cm}

{\bf 2. Rigorous definitions of the secondary and double symmetries}

\vspace{0.2 cm}

{\bf Definition 1.} Assume that there are a symmetry group $G$ of some field 
theory and two its representations $T$ and $S$. Let $\theta = \{\theta_{a}\}$ 
be a vector in the representation space of $T$, $\Psi(x)$ be any field vector 
in the representation space of $S$, and let $D^{a}$ be such operators that 
the field $\Psi'(x)$, obtaned by the transformation 
\begin{equation}
\Psi' (x) = \exp (-iD^{a}\theta_{a} )\Psi (x),
\end{equation}
belongs again to the representation space of $S$, i.e. for any $g \in G$
\begin{equation}
\exp (-iD^{b}(T(g)\theta)_{b}) S(g) \Psi (x) = S(g) \Psi '(x).
\end{equation}
Then the transformations (1) and their products will be called secondary
symmetry transformations produced by the representation $T$ of the group $G$.

The relations (1) and (2) allows to give a more general definition.

{\bf Definition 2.} Let the group ${\cal G}_{T}$ contain the subgroup $G$ 
and the invariant subgroup $H_{T}$, with $G \neq {\cal G}_{T}$,
$H_{T} \neq {\cal G}_{T}$, and ${\cal G}_{T} = H_{T} \circ G$. If any element 
$h_{T} \in H_{T}$ can be written in the form
\begin{equation}
h_{T} = h(\theta_{1})  h(\theta_{2}) \ldots h(\theta_{n}), \hspace{0.5 cm}
n \in \{1, 2, \ldots \}, 
\end{equation}
where $\theta_{i}$ $(i = 1, 2, \ldots, n)$ is some vector of the
representation space of $T$ of the group $G$, and if for any element $g \in G$
\begin{equation}
g h(\theta) g^{-1} = h(T(g)\theta ),
\end{equation}
then the group $G$  will be called the primary symmetry group, and the groups 
$H_{T}$ and ${\cal G}_{T}$ will be respectively called the secondary and 
double symmetry groups produced by the representation $T$ of the group $G$.

It follows from the relations (1) and (2) that the operators  $D^{a}$ must
satisfy the condition
\begin{equation}
D^{a} = S^{-1}(g) D^{b} S(g) [T(g)]_{b}{}^{a},
\end{equation}
i.e., as one uses to say, such operators $D^{a}$ transform as the
representation $T$ of the group $G$.

If $G$ is a Lie group, then within some vicinity $U$ of the unity
element $e$ of the group $G$ the transformation operators $S(g)$ and $T(g)$ 
can be written as
\begin{equation}
S(g) = \exp (-iL^{j}\epsilon_{j}),
\end{equation}
\begin{equation}
T(g) = \exp (-iM^{j}\epsilon_{j}),
\end{equation}
where $L^{j}$ and $M^{j}$ are generators of the group $G$ of the 
representations $S$ and $T$, $\epsilon_{j} = \epsilon_{j}(g)$ are group 
parameters corresponding to the given element $g \in U$.

By using Eqs. (6) and (7), condition (5) takes the form
\begin{equation}
[L^{j}, D^{a}] = D^{b} (M^{j})_{b}{}^{a}.
\end{equation}

Expanding the exponent of Eq. (1) in a power series, we see that the first 
term $(\Psi)$ transforms as $S$ and the second term $(\theta \cdot \Psi)$ 
does as a direct product $T \otimes S$. The following statement is getting
obvious. Nontrivial ($D^{a}\not=0$) secondary symmetry transformations (1) 
exist if and only if among irreducible representations of the direct product 
$T \otimes S$ there is at least one which belongs to the representation $S$. 
If the operators $D^{a}$ are matrix ones, then their elements are nothing but
the unnormalized Clebsch-Gordon coefficients for relevant irreducible 
representations of the group $G$ to be found from the relation (8). Namely, 
let $\omega_{t}$, $\omega_{p}$, and $\omega_{q}$ denote three irreducible 
representations of the group $G$, such that $\omega_{t} \in T$, 
$\omega_{p} \in S$, and $\omega_{q} \in T \otimes S$, and let  $\alpha_{t}$, 
$\alpha_{p}$ and $\alpha_{q}$ be a set of indices that characterize vectors 
in the spaces of relevant irreducible representations. Then we have the
well-known relations
\begin{equation}
D^{\omega_{t}\alpha_{t}}_{\omega_{q}\alpha_{q},\omega_{p}\alpha_{p}} = 0,
\hspace{0.5 cm} {\rm if} \hspace{0.3cm} \omega_{q} \not\in S,
\end{equation}
\begin{equation}
D^{\omega_{t}\alpha_{t}}_{\omega_{q}\alpha_{q},\omega_{p}\alpha_{p}} =
d^{\omega_{t}}_{\omega_{q}\omega_{p}} 
(\omega_{t}\alpha_{t}\omega_{p}\alpha_{p}|\omega_{q}\alpha_{q}), 
\hspace{0.5 cm} {\rm if} \hspace{0.3 cm} \omega_{q} \in S,
\end{equation}
where $(\omega_{t}\alpha_{t}\omega_{p}\alpha_{p}|\omega_{q}\alpha_{q})$ are
the properly normalized Clebsch-Gordon coefficients, 
$d^{\omega_{t}}_{\omega_{q}\omega_{p}}$ are arbitrary quantities independent 
on the indices $\alpha_{t}$, $\alpha_{p}$ and $\alpha_{q}$.

\vspace{0.3 cm}

{\bf 3. Two ways to construct double-symmstric field theories}

\vspace{0.2 cm}

Structure of the groups $H_{T}$ and $\cal{G}_{T}$, produced by the
representation $T$ of the group $G$, depends on the representation $S$ of the
group $G$ and on numerical values of the quantities determining the operators
$D^{a}$.

If the secondary symmetry is produced by the adjoint representation of the
group $G$ and the operators $D^{a}$ of Eq. (1) coinside with the group
generators, then obviously the group $H_{T}$ is locally isomorphic to the
group $G$. In such a case we shall say that the double symmetry is degenerated
one.

In the general case there are at least two ways to complete the construction 
of the group $\cal{G}_{T}$ and to construct the double-symmetric field
theories, i.e. the theories whose Lagrangians are invariant under
transformations of the group $\cal{G}_{T}$.

The first way is to close the algebra of the operators $D^{a}$ starting from
those or other considerations on the fields in question and their 
interactions. Possessing the Lie algebra of the group $\cal{G}_{T}$ one can 
find then its representations and allowed double-symmetric theories. One 
proceeds so in the supersymmetry theory [1-3].

Desribing the second way for some fixed representation $T$, we consider, for 
simplicity, the matrix realization of the operators $D^{a}$. First we choose 
some class of representations of the group $G$ which seem to be admissible 
for the field $\Psi (x)$. Some set of arbitrary quantities 
$d^{\omega_{t}}_{\omega_{q}\omega_{p}}$ determining the operators $D^{a}$ 
through Eq. (10) corresponds to each representation of this class. It is 
required the Lagrangian of the considered theory to be invariant under 
transformations of the group $G$ as well as under global transformations (1). 
Then, obviously, it will be invariant under any transformations of the groups 
$H_{T}$ and $\cal{G}_{T}$. These requirements can lead to some selection of 
the representations of the considered class, and restrict the arbitrariness 
of quantities $d^{\omega_{t}}_{\omega_{q}\omega_{p}}$ and the arbitrariness 
of Lagrangian constants allowed by the invariance in respect to the group $G$.
If the arbitrariness of quantities $d^{\omega_{t}}_{\omega_{q}\omega_{p}}$ is 
removed completely (up to a common constant) for each of selected 
representations of the group $G$, then the structure of the group $H_{T}$ 
becomes automatically fully defined, but, generally speaking, it is unlike 
for different representations of the group $G$. If it is not so, only then a 
question arises on closing the algebra of the operators $D^{a}$ taking into 
account the obtained restrictions. Thus, on this way the first place is taken 
by constraints established by the double symmetry produced by the 
representation $T$ of the group $G$, and the problem on the Lie algebra of 
the group $\cal{G}_{T}$ is resolved automatically or moved onto the last 
place. It seems to be reasonable to demonstrate the efficiency of the 
described way with the help of a nontrivial example. Let us do this.

\vspace{0.3 cm}

{\bf 4. An example of constructing double-symmetric field theory: 
the efficiency of selecting representations and removing ambiguities}

\vspace{0.2 cm}

Consider the relativistically invariant \footnote{Here, as well as in the
monograph [4], the relativistic invariance means, in modern titles and
notations [5], an invariance in respect to the orthochronous Lorentz group 
$L^{\uparrow}$ generated by the proper Lorentz group $L^{\uparrow}_{+}$ and 
spatial reflection $P$.} Lagrangian of the general type [4] for some free 
fermionic field $\Psi (x)$
\begin{equation}
{\cal L}_{0} = \frac{i}{2}[(\partial_{\mu} \Psi, L^{\mu} \Psi ) -
(\Psi, L^{\mu} \partial_{\mu} \Psi )] - \kappa (\Psi, \Psi ),
\end{equation}
where $(\Psi_{1}, \Psi_{2})$ is a relativistically invariant belinear form,
$L^{\mu}$ are matrix operators, and $\kappa$ is a constant.

Require the Lagrangian (11) to be invariant under the global secondary
symmetry transformations produced by a polar 4-vector of the group
$L^{\uparrow}$
\begin{equation}
\Psi '(x) = \exp (-i D^{\mu} \theta^{\mu}) \Psi (x),
\end{equation}
where $D^{\mu}$ are matrix operators. This requirement will be fulfiled if
\begin{equation}
[ L^{\mu}, D^{\nu} ] = 0,
\end{equation}
\begin{equation}
(D^{0} \Psi_{1}, \Psi_{2}) = (\Psi_{1}, D^{0} \Psi_{2}).
\end{equation}

The condition (5) for the operators $D^{\mu}$ and the condition, which the 
operators $L^{\mu}$ of the Lagrangian (11) obey, are equivalent. Therefore 
the matrix elements of both operators $L^{\mu}$ and $D^{\mu}$ are given by
relations of the type (9-10). In the monograph [4] there are given their 
explicit forms. We shall use notations \footnote{In these notations all 
irreducible representations of the group $L^{\uparrow}_{+}$ are elegantly 
described: finite- and infinite-dimensional ones. Transition to the often 
used notations $(j_{1}, j_{2})$, which are connected with the group 
$SO(4) = SO(3) \otimes SO(3)$ and describe only finite-dimensional 
irreducible representations, is the following: $j_{1} = (l_{1}+l_{0}-1)/2$, 
$j_{2} = (l_{1}-l_{0}-1)/2$.} $\tau = (l_{0}, l_{1})$ of Ref. [4] for 
irreducible representations of the proper Lorentz group and denote the 
arbitrary quantities of Eq. (10), characterizing the operators $L^{\mu}$ and 
$D^{\mu}$, as $c_{\tau' \tau}$ and $d_{\tau' \tau}$, respectively.

Let the representation $S$ of the group $L{\uparrow}$, as which the field
$\Psi (x)$ of the Lagrangian (11) transforms, be decomposible into a finite
or infinite direct sum of irreducible finite-dimentional representations with 
half-integer spins, and let the multiplicity of any of these irreducible 
representations do not exeed 1. Then the Lagrangian (11) will be invariant 
under global transformations (12) if and only if $S$ is one of the 
infinite-dimentional representations $S^{k_{1}}$
\begin{equation}
S^{k_{1}} = \sum _{n_{1}=0}^{\infty} \sum_{k_{0}=-k_{1}+1}^{k_{1}-1} \oplus
(k_{0}, k_{1}+n_{1}), \hspace{0.5cm} k_{1} = 3/2, 5/2, \ldots
\end{equation}
or is the representation $S^{F}$ containing all irreducible representations of
the group $L^{\uparrow}_{+}$ with half-integer spins.

For each of these representations all the quantities $|c_{\tau' \tau}|$ and
$|d_{\tau' \tau}|$ have, up to their common constants, fully defined values.
For any of the representations (15) we get in result
\begin{equation}
[ D^{\mu}, D^{\nu} ] = 0,
\end{equation}
i.e. the secondary symmetry group $H_{T}$ is the four-parameter Abelian
group. Notice that, although in this case the Lie algebra of the double
symmetry group ${\cal G}_{T}$ coinsides with the Lie algebra of the
Poincar\`{e} group, the operators $D^{\mu}$ cannot be identified with the 
translation operators $P^{\mu}$ because $D^{\mu}$ act only on spin variables 
of a field but $P^{\mu}$ do not act on them.

The mass spectra corresponding to each of the obtained Lagrangians (11) are
infinetely degenerated in spin and continious. In order to eliminate this
degeneration it is necessary to break the secondary symmetry spontanously
keeping the orthochronous Lorentz group symmetry.

\vspace{0.3 cm}

{\bf 5. On a perspective to use the double symmetry}

\vspace{0.2 cm}

The considered example of the double symmetry, playing here only the
methodological role, is used by us as one of elements for constructing a
theory with the fields which transform as orthochronous Lorentz group 
representations decomposible into  an infinite direct sum of 
finite-dimentional irreducible representations \footnote{Two our articles on 
the theory of infinite-component fields with the double symmetry, produced by 
the polar and axial 4-vectors of the group $L^{\uparrow}$, are preparing for
publication}. Former the attemps to analize the theory with such class 
fields were not undertaken because of the infinite number of arbitrary 
constants and the absence of criterions for removing this arbitrariness. 
Carried out sometime ago investigations of the infinite-component field 
theories, which were assigned to an alternative describtion of hadrons as 
composite particles, were based on those representations of the Lorentz group 
that could be decomposed into a finite direct sum of infinite-dimentional 
irreducible representations or on a special representation of the group 
$SO(4,1)$ or $SO(4,2)$. It was established, however, that such theories 
possesses some properties (peculiarities of mass spectra, nonlocality, 
violation of $CPT$-invariance and the connection between spin and statistics) 
which are not admissible for particle physics. Problems, covered by these 
investigations as well as a sufficiently full list of relevant papers, can be 
found in the monograph [5].

\vspace{0.3 cm}

{\bf 6. Existing undegenerated double symmetry: supersymmetry}

\vspace{0.2 cm}

Notice now that sypersymmetry both in the $x$-space and in the superspace can 
be considered as a double symmetry produced by the bispinor representation
$T$ of the proper Lorentz group. An element $h(\theta )$ of the corresponding
secondary symmetry group $H_{T}$ has a form [2]
\begin{equation}
h(\theta ) = \exp (iQ_{\alpha} \theta^{\alpha} + 
i \overline{Q}_{\dot{\alpha}} \overline{\theta}^{\dot{\alpha}}),
\end{equation}
the parameters $\theta^{\alpha}$ and $\overline{\theta}^{\dot{\alpha}}$
being anticommutating elements of the Grassmann algebra and belonging to the
representation spaces of $(1/2, 0)$ and $(0, 1/2)$ of the proper Lorentz
group, respectively (in the notations connected with the group $SO(4)$). A
fulfilment of the relation (8), as a criterion of the secondary symmetry, was
already required in respect to the bispinor operators $Q_{\alpha}$ and
$\overline{Q}_{\dot{\alpha}}$ in the first paper on the supersymmetry algebra
[1]. In the supersymmetry theory the operators $Q_{\alpha}$ and
$\overline{Q}_{\dot{\alpha}}$ are given in the form of sum of terms containing
a first or zero power of the differential operator $\partial_{\mu}$ (cf.
[2-3]). Such realization of the operators $Q_{\alpha}$ and 
$\overline{Q}_{\dot{\alpha}}$ leads to that the secondary symmetry group 
$H_{T}$, generated by the elements (17), contains the space-time translation
group, and the double symmetry group $\cal{G}_{T}$ contains the Poincar\`{e}
group. If one chose a matrix realization for the operators $Q_{\alpha}$ and
$\overline{Q}_{\dot{\alpha}}$, then we would come out of the supersymmetry
theory standards.

Let for a given representation $S$ of the group $L^{\uparrow}_{+}$ the 
relation (8) for the spinor operators $Q_{\alpha}$ in some its realization 
keep arbitrary $N$ constants. Then there are $N$ linearly independent spinor 
operators $Q_{\alpha}^{1}$, $Q_{\alpha}^{2}$, $\ldots$, $Q_{\alpha}^{N}$, in 
this realization, satisfying the relation (8). In this case, on one hand, one 
can introduce a secondary symmetry produced by the $N$-multiple bispinor
representation of the proper Lorentz group, and, on another hand, treat the
numbers $1, 2, \ldots , N$ of the spinor operators as an index related to some
inner symmetry. If as well the operators $Q_{\alpha}^{j}$ and 
$\overline{Q}_{\dot{\alpha}}^{j}$ ($j = 1, \ldots , N$) keep a linear
dependence on the operator $\partial_{\mu}$, then the relevant double symmetry
is an extended supersymmetry (cf. [2-3]).

\vspace{0.3 cm}

{\bf 7. Existing undegenerated double symmetry: the $\sigmab$-model symmetry}

\vspace{0.2 cm}

Another example of an undegenerated double symmetry, used in particle physics,
is the $\sigma$-model symmetry, the most accurately described by Gell-Mann and
Levy [6]. Infinitesimal transformations in the $\sigma$-model have the forms
\begin{equation}
N' = (1 - \frac{i}{2} \gamma^{5} {\taub} {\thetab})N,
\end{equation}
\begin{equation}
{\pib}' = {\pib} + i {\thetab} \sigma ,
\end{equation}
\begin{equation}
\sigma' = \sigma - i {\thetab} {\pib} ,
\end{equation}
where the field $N$ is a nucleon isodoublet, the field ${\pib}$ is a
pseudoscalar isotriplet, the field $\sigma$ is a scalar isosinglet, and the
transformation parameter ${\thetab}$ is a pseudoscalar isotriplet.

The listed transformation properties of the field and the parameter 
${\thetab}$ allow to state that the transformations (18) and (19-20) are
transformations of the secondary symmetry produced by the representation 
$T$ = ({\it isotriplet, pseudoscalar}) or, that is equivalent, by the 
representation $T$ = ({\it isotriplet, scalar}) $\oplus$ ({\it isotriplet, 
pseudoscalar}) of the group $G = SU(2) \otimes L^{\uparrow}$. We wish 
especially emphasize that, firstly, transformations (18) and (19-20) do not 
violate the spatial reflection symmetry, and so the group $G$ contains the 
orthochronous Lorentz group, and, secondly, the parity of fields, involved in 
the transformations (19-20), are necessarily different due to a pseudoscalar 
character of the paremeter ${\thetab}$. $G_{T} = SU(2)_{L} \otimes SU(2)_{R}$ 
is a group of the secondary symmetry generated by the transformations (18) 
and (19-20), the parameters of one of the group $SU(2)$ being given by sum of 
the space scalar and pseudoscalar, and the other group parameters being given 
by their difference.

\vspace{0.3 cm}

{\bf 8. Dual status of the group} {\boldmath $SU(2)_{L} \otimes SU(2)_{R}$}

\vspace{0.2 cm}

In numerous works including the pioneer papers by Schwinger [7], G\"{u}rsey
[8] and Touschek [9], which used the chiral symmetry group $SU(N)_{L} 
\otimes SU(N)_{R}$ or the group $U(1)_{A}$, it is difficult, if it's really
possible, to find any commentaries on transformation properties of these group
parameters in respect to the spatial reflection. As a rule it follows from the
contents of the papers that the transformations of the group 
$SU(N)_{L} \otimes SU(N)_{R}$ or the group $U(1)_{A}$ do not violate
$P$-symmetry. This rule, however, is broken in the left-right symmetric model 
of electroweak interactions [10-12] and in the unifited models of strong and 
electroweak interactions including the first one as its element. Actually, in 
the papers [10-12] there are not at all indications that any component of one 
or another Higgs multiplet is the space pseudoscalar. We are forced to admit 
that all components of all Higgs multiplets, and, consequently, all 
parameters of the used group $SU(2)_{L} \otimes SU(2)_{R}$ are space scalars. 
Transformations of this group, generally speaking, violate the $P$-symmetry. 
If, for example, the initial state of a spinor field possesses a definite 
parity, then the state, obtained due to a transformation of the type (18), 
already does not possess it. The group in question itself does not establish 
any relation between coupling constants of two $W$-bosons to fermions. 
Therefore in the left-right symmetric model one introduces some discrete 
symmetry which transforms $SU(2)_{L}$ to $SU(2)_{R}$ but it is not identified 
with the spatial reflection. So one can state that the group 
$SU(2)_{L} \otimes SU(2)_{R}$ of this model is a secondary symmetry group 
produced by 2-multiple scalar isotriplet of the group 
$G = SU(2) \otimes L^{\uparrow}_{+} \otimes$ ({\it A discrete symmetry
group}), this discrete symmetry being not fully defined.

\vspace{0.3 cm}

{\bf 9. On {\boldmath $P$}-properties of the physical vacuum and the
gauge fields of electroweak interactions}

\vspace{0.2 cm}

Clarify now principal points related to the $P$-symmetry and its violation in
electroweak interactions, if one corrects the left-right symmetric model so
that it would be $P$-invariant before the spontaneous symmetry breaking. Note
at once, that this correction does not affect values of the cross-sections 
and the decay probabilities.

Initial $P$-invariance and its observed violation is ensured by the local
double symmetry produced by the representation $T$ = ({\it isotriplet, 
scalar}) $\oplus$ ({\it isotriplet, pseudoscalar}) $\oplus$ ({\it isosinglet, 
scalar}) $\equiv$ $(1, s) \oplus (1, p) \oplus (0, s)$ of the group 
$G = SU(2) \otimes L^{\uparrow}$. Corresponding parameters of the secondary 
symmetry transformations and the gauge fields will be denoted as 
${\thetab}^{1s} = \{ \theta^{1s}_{j} \}$,
${\thetab}^{1p} = \{ \theta^{1p}_{j} \}$, $\theta^{0s}$; 
${\bf B}^{1V}_{\mu} = \{ B^{1V}_{j \mu} \}$, 
${\bf B}^{1A}_{\mu} = \{ B^{1A}_{j \mu} \}$, $B^{0V}_{\mu}$ ($j = 1, 2, 3$;
the indices $V$ and $A$ mean the polar and axial 4-vectors, respectively).

In the fermionic sector we restrict ourselves by isodoublet consisting of the
fields of electronic neutrino $\nu_{e}$ and electron $e$. Its seconary 
symmetry transformations, as well as all anothers, are written in the form,
explicitly satisfying the conditions of Definition 1
\begin{equation}
\psi' = \exp \left( -\frac{i}{2}{\taub}{\thetab}^{1s} - \frac{i}{2} \gamma^{5}
{\taub}{\thetab}^{1p} + \frac{i}{2} \theta^{0s} \right) \psi,
\end{equation}
with $\psi^{T} = (\nu_{e}, \; e)$.

Checking that the transformations (21) constitute a group we introduce the
above mentioned gauge fields with such phases that in the Lagrangian of their
interaction with leptonic field $\psi$
\begin{equation}
{\cal L}_{int} = -\frac{1}{2\sqrt{2}}\overline{\psi} \left( g_{1V}\gamma^{\mu}
{\taub} {\bf B}^{1V}_{\mu} + g_{1A} \gamma^{\mu} \gamma^{5}
{\taub} {\bf B}^{1A}_{\mu} - g_{0} \gamma^{\mu} B^{0V}_{\mu} \right) \psi 
\end{equation}
the coupling constants $g_{1V}$, $g_{1A}$ and $g_{0}$ are positive. The
formulae (21) and (22) give also a knowledge of covariant derivatives
corresponding those or others gauge transformations of the secondary symmetry
group. 
 
The most general form of the global secondary symmetry transformations of the
fields ${\bf B}^{1V}_{\mu}$ and ${\bf B}^{1A}_{\mu}$ is
\begin{equation}
{{\bf B}^{1V}_{\mu}\choose {\bf B}^{1A}_{\mu}}' = \exp \left[ -i\left( \matrix
{a_{1}{\bf t}&0\cr 0&a_{2}{\bf t}} \right) {\thetab}^{1s} - i\left( \matrix
{0&b_{1}{\bf t}\cr b_{2}{\bf t}&0} \right) {\thetab}^{1p} \right] 
{{\bf B}^{1V}_{\mu}\choose {\bf B}^{1A}_{\mu}}, 
\end{equation}
where ${\bf t} = \{ t_{j} \}$ ($j = 1, 2, 3$) are the generators of adjoint 
representation of the group $SU(2)$, and $a_{1}$, $a_{2}$, $b_{1}$ and 
$b_{2}$ are arbitrary constants. A requirement the transformation (23) to be 
orthogonal and the Lagrangian (22) to be invariant under the global
transformations (21) and (23) is filfiled if and only if
$a_{1} = a_{2} = b_{1} = b_{2} = 1$ and $g_{1V} = g_{1A} \equiv g$.

We introduce the Higgs field $\Phi$ consisting of scalar $\phi^{\frac{1}{2}s}$
and pseudoscalar $\phi^{\frac{1}{2}p}$ isodoublets and taking some vacuum 
expectation values of its neutral components with isospin projection $-1/2$: 
$<\phi^{\frac{1}{2}s}_{-1/2}> = v_{s}$, $<\phi^{\frac{1}{2}p}_{-1/2}> = 
v_{p}$. Relative phases of the fields $\phi^{\frac{1}{2}s}$ and
$\phi^{\frac{1}{2}p}$ are fixed so that the secondary symmetry transformations
have the form
\begin{equation}
{\phi^{\frac{1}{2}s}\choose\phi^{\frac{1}{2}p}}' = \exp \left[
-\frac{i}{2} \left( \matrix{{\taub}&0\cr 0&{\taub}} \right)
{\thetab}^{1s} - \frac{i}{2} \left( \matrix{0&{\taub}\cr
{\taub}&0} \right) {\thetab}^{1p} - \frac{i}{2} \theta^{0s}
\right] {\phi^{\frac{1}{2}s}\choose\phi^{\frac{1}{2}p}}.
\end{equation}

The Higgs fields $\phi^{\frac{1}{2}s}$ and $\phi^{\frac{1}{2}p}$ are
sufficient to reproduce all results of the Weinberg-Salam model in the region
of existing energies, except for the generation of fermionic masses. It's 
just needed that the relation $|v_{s}-v_{p}| <\!< |v_{s}+v_{p}|$ would be 
fulfiled and the vacuum expectation value of any other field of Higgs's type 
would be much less then $|v_{s}-v_{p}|$. Concerning fermionic masses one can 
suggest up to some moment that they appear as a result of 
interaction of fermions with Higgs fields constituting the multiplets 
$\{ \phib^{1s}, \phi^{0p} \}$ and $\{ \phib^{1p}, \phi^{0s} \}$, whose 
transformations are generated by the elements of the type (19-20).

In further formulae we neglect the vaccum expectation values  af all fields 
apart from $\phi^{\frac{1}{2}s}$ and $\phi^{\frac{1}{2}p}$. From the 
Lagrangian ${\cal L}_{\Phi} = |{\cal D}_{\mu} \Phi |^{2}$, where 
${\cal D}_{\mu}$ is a covariant derivative corresponding to the gauge 
transformations (24), we get that the electromagtetic field $A_{\mu}$, the 
fields of light $W^{(1)\pm}_{\mu}$, $Z^{(1)}_{\mu}$ and heavy 
$W^{(2)\pm}_{\mu}$, $Z^{(2)}_{\mu}$ intermediate bosons are described by the 
following relations
\begin{equation}
A_{\mu} = \frac{1}{\sqrt{g^{2} + g^{2}_{0}}}(g_{0}B^{1V}_{3\mu}
+ gB^{0V}_{\mu}),
\end{equation}
\begin{equation}
W^{(i)\pm}_{\mu} = \frac{1}{2} [(B^{1V}_{1\mu} - \eta_{i} B^{1A}_{1\mu})
\mp i(B^{1V}_{2\mu} - \eta_{i} B^{1A}_{2\mu})],
\end{equation}
\begin{equation}
Z^{(i)}_{\mu} = \frac{\alpha_{i}}{\sqrt{g^{2}+g^{2}_{0}}}
(gB^{1V}_{3\mu} - g_{0}B^{0V}_{\mu}) - \beta_{i} B^{1A}_{3\mu},
\end{equation}
\begin{equation}
m^{2}_{W^{(i)}} = \frac{g^{2}}{4} |v_{s} - \eta_{i} v_{p} |^{2},
\end{equation}
\begin{equation}
m^{2}_{Z^{(i)}} = \frac{|v_{s}|^{2} + |v_{p}|^{2}}{4} \left( g^{2} +
\frac{g^{2}_{0}}{2} - \eta_{i} \frac{g^{2}_{0} \sqrt{\gamma^{2} + 1}}
{2 \gamma} \right),
\end{equation}
where
$$\eta_{1} = 1, \hspace{0.5cm} \eta_{2} = -1, \hspace{0.5cm}
\gamma = \frac{g^{2}_{0}(|v_{s}|^{2} + |v_{p}|^{2})}
{2g\sqrt{g^{2}+g^{2}_{0}}(v_{s}v^{*}_{p}+v^{*}_{s}v_{p})},$$
\begin{equation}
\alpha_{i} = \sqrt{\frac{1}{2} \left( 1-\eta_{i} \frac{\gamma}
{\sqrt{\gamma^{2}+1}} \right)},  \hspace{0.5cm}
\beta_{i} = \eta_{i} \sqrt{\frac{1}{2} \left( 1+\eta_{i} \frac{\gamma}
{\sqrt{\gamma^{2}+1}} \right)}.
\end{equation}

From Eqs. (22) and (25-27) we get
$${\cal L}_{int} = e_{0} \overline{e} \gamma^{\mu} e A_{\mu} - \frac{1}
{2\sqrt{2}} \sum^{2}_{i=1} [ g\overline{\nu}_{e} \gamma^{\mu} (1-
\eta_{i} \gamma^{5} ) e W^{(i)+}_{\mu} + {\rm H.c.}$$
\begin{equation}
+ \overline{\nu}_{e} (\alpha_{i} \sqrt{g^{2}+g^{2}_{0}} \gamma^{\mu} -
\beta_{i} g \gamma^{\mu} \gamma^{5} )\nu_{e} Z^{(i)}_{\mu} +
\overline{e} (-\alpha_{i} \frac{g^{2}-g^{2}_{0}}{\sqrt{g^{2}+g^{2}_{0}}}
\gamma^{\mu} +\beta_{i} g \gamma^{\mu} \gamma^{5} )e Z^{(i)}_{\mu} ].
\end{equation}

It follows from Eqs. (28) and (31) that the observed domination of the 
left-hand week charged current is possible if and only if $v_{s} \neq 0$, 
$v_{p} \neq 0$ and $\arg (v_{p}/v_{s}) \neq \pm \pi /2$. This means that 
{\bf the physical vacuum does not possess of a definite 
{\boldmath $P$}-parity} because $Pv_{s} = v_{s}$, $Pv_{p} = - v_{p}$. This is 
the first principal point which has not been clear in the left-right 
symmetric model. It's interesing, that at the time when the spontaneous 
breaking of the gauge symmetry and a related to it understanding of physical 
vacuum was not been formulated yet, Nambu and Jona-Lasinio noted in their 
paper [13], "that the $\gamma^{5}$ transformation changes the parity of the 
vacuum which will be in general a superposition of states of opposite 
parities".

It follows from Eqs. (26-27) that {\bf the fields of all intermediate bosons
constitute a superposition of polar and axial 4-vectors, these vectors having 
an equal weight in the fields of {\boldmath $W$}-bosons}. This is just the
second principal point which has not been clear in the left-right symmetric
model, though, on our opinion, it has the same powerful significance as the
form of weak currents.

If we anywhere considered the intermediate boson masses and the coupling
constants of $Z$-bosons to fermions as functions of the field values $v_{s}$ 
and $v_{p}$, then we would get from Eqs. (26)-(31) that under the spatial 
reflection the $Z^{(1)}$-boson coupling constants turn into the 
$Z^{(2)}$-boson coupling constants and vise versa, as well as
\begin{equation}
P m^{2}_{W^{(1)}} = m^{2}_{W^{(2)}}, \hspace{0.2cm}
P m^{2}_{Z^{(1)}} = m^{2}_{Z^{(2)}}, \hspace{0.2cm}
P W^{(1)\pm}_{\mu} = (-1)^{\delta_{\mu}}  W^{(2)\pm}_{\mu}, \hspace{0.2cm}
P Z^{(1)\pm}_{\mu} = (-1)^{\delta_{\mu}}  Z^{(2)\pm}_{\mu},
\end{equation}
where $\delta_{\mu} = 0$ if $\mu = 0$, and $\delta_{\mu} = 1$ if 
$\mu = 1,2,3$. This would cause an invariance of the Lagrangian (31).

Thus, the corrected version of the left-right symmetric model of electroweak 
interactions leads to the logically completed interpretation of the
$P$-invariance violation, and reveals a full analogy between transformation
properties in respect to the spatial reflection of week currents and 
corresponding gauge fields of the intermediate bosons.

\vspace{0.3 cm}

{\bf Ackowledgements}

\vspace{0.2 cm}

I am very grateful to V.I. Savrin, I.P. Volobuev and N.P. Yudin for useful
discussions and their support of my work.

\end{document}